Relativistic five-quark equations and u, d- pentaquark spectroscopy.

Gerasyuta S.M.[1,2,3], Kochkin V.I.[1]

1. Department of Theoretical Physics, St. Petersburg State University, 198904, St. Petersburg, Russia.
2. Department of Physics, LTA, 194021 St. Petersburg, Russia.
3. Forschungzentrum Julich, Institut fur Kernphysik (Theorie) D-52425 Julich, Germany.

Abstract.

The relativistic five-quark equations are found in the framework of the dispersion relation technique. The five-quark amplitudes for the low-lying pentaquarks including $u,d$ quarks are calculated. The poles of the five-quark amplitudes determine the masses of the lowest pentaquarks. The calculation of pentaquark amplitudes estimates the contributions of four subamplitudes. The main contributions to the pentaquark amplitude are determined by the subamplitudes, which include the meson states $M$.

I. Introduction

Recently the LEPS collaboration at Spring-8 presented evidence for the existence of a narrow baryon resonance with strangeness S = +1 [1]. In the following four other collaboration from different laboratories announced the observation of similar structure in their experiments [2-5]. The observed structure was immediately brought into connection with exotic pentaquark state called $\theta^+$ whose existence had been proposed since long time in the context of different quark models. Speciffically, the resonance parameters with a peak position around 1540 MeV and a width around 20 MeV, extracted from these experiments, lie convincingly close to the theoretical prediction based on the chiral quark-soliton model of Diakonov et al [6], who had proposed the existence of a $\theta^+$ state with amass around 1530 MeV and a width of around 15 MeV. Due to its quantum numbers, S = +1, $I = 0$ and $J^P = \frac{1}{2}^+$, their $\theta^+$ state can only decay (hadronically) into the $K^+n$ or $K^0 p$ channels.

Capstick, Page and Roberts hypothesized that the $\theta^+$ is interpreted as a state containing a dominant pentaquark Fock-state component $uudd\bar{s}$ decaying to $K^+n$, in contrast to its interpretation as a chiral soliton. In this pure multiquark picture $\theta^+$ has isospin, spin and parity different from those predicted by the Skyrme model [6], e.g. $\theta^+$ can be isotensor with $J^P = \frac{1}{2}^-, \frac{3}{2}^-, \frac{5}{2}^-$ [7].

Jaffe and Wilczek suggested that the observed $\theta^+$ state could be composed of an antistrange quark and two highly correlated up and down pairs arising from strong color-spin correlation force [8]. The resulting $J^P$ of $\theta^+$ state is $\frac{1}{2}^+$.

In our paper [9] the relativistic generalization of the four-body Faddeev - Yakubovsky type equations are represented in the form of dispersion relations of the two-body subenergy. We investigated the relativistic scattering four-body amplitudes of the constituent quarks of two flavors (u, d). The poles of these amplitudes determine the masses of the lowest hybrid mesons. The constituent quark is color triplet and quark amplitudes obey global color symmetry. We used the results of the bootstrap quark model [10] and introduced the $q\bar{q}$ – state in the color octet channel with $J^{PC} = 1^{--}$ and isospin $I = 0$. This bound state is identified as a constituent gluon. In our model we take into account the color state with $J^{PC} = 1^{--}$ and isospin $I = 1$, which determines the hybrid state with the constituent gluon. In addition, $q\bar{q}q\bar{q}$ states are also predicted.

We derived the mixing of the hybrid and $q\bar{q}q\bar{q}$ states. This state was called the hybrid meson. The mass spectrum of lowest hybrid mesons with isospin $I = 1$ both with exotic quantum numbers (non-$q\bar{q}$) $J^{PC} = 1^{-+}$, $0^{--}$ and ordinary quantum numbers $J^{PC} = 0^{++}$, $1^{++}$, $2^{++}$, $0^{-+}$, $1^{--}$ was calculated. The important result of this model is the calculation of hybrid meson amplitudes, which contain the contribution of two subamplitudes: four-quark amplitude and hybrid amplitude. The main contribution corresponds to the four-quark amplitude. The hybrid amplitude gives rise to only less than 40 % of the hybrid meson contribution.

In our previous paper [11] the relativistic generalization of five-quark equations (like Faddeev – Yakubovsky approach) are constructed in the form of the dispersion relation. The five-quark amplitudes for the low-lying hybrid baryons contain only light quarks and are calculated under the condition that flavor $SU(3)_f$ symmetry holds. In should be noted, that the calculated masses of low-lying hybrid baryons agree with data [12] and with the results obtained in the flux-tube model [13].

In our relativistic quark model with four-fermion interaction the octet color $q\bar{q}$ bound state was found, which corresponds to the constituent gluon $G$ with a mass $M_G = 0.67$ GeV [9]. This approach is similar to the large $N_c$ limit [14-16]. In diquark channel we have the diquark level D with $J^P = 0^+$ and the mass $m_{u,d} = 0.72$ GeV (in the color state $\bar{3}_c$). The diquark state with $J^P = 1^+$ in color state $\bar{3}_c$ also has an attractive interaction, but smaller than that of the diquark with $J^P = 0^+$; therefore there is only the correlation of quarks, not a bound state [10].

The calculated five-quark amplitude consists of four subamplitudes: $qDG$, $qqqG$, $D\bar{q}D$ and $qq\bar{q}D$, where $D$ and $G$ are the diquark state and exited constituent gluon state respectively. The main contributions to the hybrid baryon amplitude are determined by the subamplitudes, which include the exited gluon states.

The present paper is devoted to the construction of relativistic five-quark equations for two flavors (u, d) pentaquarks. The five-quark amplitudes for the lowest pentaquarks contain only light quarks. The poles of these amplitudes determine the masses of the $udud\bar{u}$ and $uuuu\bar{u}$ pentaquarks. The constituent quark is the color triplet and the quark amplitudes obey the global color symmetry. The interesting result of this model is the calculation of pentaquark amplitudes which contain the contribution of four subamplitudes: molecular subamplitude $BM$, $D\bar{q}D$ subamplitude, $Mqqq$ subamplitude and $Dqq\bar{q}$ subamplitude. Here $B$ corresponds to the lowest baryon (nucleon and $\Delta$–isobar baryon). $M$ are the low-lying mesons with the quantum numbers: $J^{PC} = 0^{++}, 1^{++}, 2^{++}, 0^{-+}, 1^{--}$ and $I_z = 0, 1$. We call the pentaquark with $J^P = \frac{1}{2}^+$ the $N$ pentaquark and the pentaquark with $J^P = \frac{3}{2}^+$ the $\Delta$ – isobar pentaquark.

The mass values of the low-lying pentaquarks are calculated (Table 1, 2). The lowest mass of $N$ pentaquark with $J^P = \frac{1}{2}^+$ is equal to M=1686 MeV. The pentaquark amplitudes take into account the contribution of four subamplitudes. The main contributions to the pentaquark amplitude are determined by subamplitudes which include the low-lying meson with $J^{PC} = 0^{++}, 1^{++}, 2^{++}, 0^{-+}, 1^{--}$.

The paper is organized as follows. After this introduction, we discuss the five-quark amplitudes which contain only light quarks (section 2). In the section 3, we report our numerical results (Tables 1, 2) and the last section is devoted to our discussion and conclusion.

In the Appendix A we give the relations, which allow to pass from the integration of the cosines of the angles to the integration of the subenergies. In the Appendix B we describe the integration contours of functions $J_1$, $J_2$, $J_3$, which are determined by the interaction of the five quarks.

## II. Pentaquark amplitudes.

We derived the relativistic five-quark equations in the framework of the dispersion relation technique. We use only planar diagrams; the other diagrams due to the rules of $1/N_c$ expansion [14-16] are neglected. The correct equations for the amplitude are obtained by taking into account all possible subamplitudes. It corresponds to the division of the complete system into subsystems with a smaller number of particles. Then one should represent a five-particle amplitude as a sum of ten subamplitudes: $A = A_{12} + A_{13} + A_{14} + A_{15} + A_{23} + A_{24} + A_{25} + A_{34} + A_{35} + A_{45}$. In our case all particles are identical, therefore we need to consider only one group of diagrams and the amplitude corresponding to them, for example $A_{12}$. The set of diagrams associated with the amplitude $A_{12}$ can be further broken down into four groups corresponding to amplitudes $A_1(s, s_{1234}, s_{12}, s_{34})$, $A_2(s, s_{1234}, s_{25}, s_{34})$, $A_3(s, s_{1234}, s_{13}, s_{134})$, $A_4(s, s_{1234}, s_{23}, s_{234})$ (Fig. 1). The antiquark is shown by the arrow and the other lines correspond to the quarks. The coefficients are determined by the permutation of quarks [17, 18].

In order to represent the subamplitudes $A_1(s, s_{1234}, s_{12}, s_{34})$, $A_2(s, s_{1234}, s_{25}, s_{34})$, $A_3(s, s_{1234}, s_{13}, s_{134})$, and $A_4(s, s_{1234}, s_{23}, s_{234})$ in the form of a dispersion relation it is necessary to define the amplitudes of quark-quark and quark-antiquark interaction $b_n(s_{ik})$. The pair quarks amplitudes $q\bar{q} \to q\bar{q}$ and $qq \to qq$ are calculated in the framework of the dispersion N/D method with the input four-fermion interaction with quantum numbers of the gluon [10]. We use the results of our relativistic quark model [10] and write down the pair quarks amplitude in the form:

$$b_n(s_{ik}) = \frac{G_n^2(s_{ik})}{1 - B_n(s_{ik})}, \quad (1)$$

$$B_n(s_{ik}) = \int_{4m^2}^{\Lambda} \frac{ds'_{ik}}{\pi} \frac{\rho_n(s'_{ik}) G_n^2(s'_{ik})}{s'_{ik} - s_{ik}}. \quad (2)$$

Here $s_{ik}$ is the two-particle subenergy squared, $s_{ijk}$ corresponds to the energy squared of particles $i$, $j$, $k$, $s_{ijkl}$ is the four-particle subenergy squared and $s$ is the system total energy squared. $G_n(s_{ik})$ are the quark-quark and quark-antiquark vertex functions (Table 3). $B_n(s_{ik})$, $\rho_n(s_{ik})$ are the Chew-Mandelstam functions with the cut – off $\Lambda$ [19] and the phase spaces respectively.

The two-particle phase space for the equal quark masses is defined as:

$$\rho_n(s_{ik}, J^{PC}) = \left(\alpha(J^{PC}, n) \frac{s_{ik}}{4m^2} + \beta(J^{PC}, n)\right) \sqrt{\frac{s_{ik} - 4m^2}{s_{ik}}},$$

The coefficients $\alpha(J^{PC},n)$ and $\beta(J^{PC},n)$ are given in Table 4.

Here n=1 corresponds to a $qq$-pair with $J^P = 0^+$ in the $\bar{3}_c$ color state, n=2 describes a $qq$-pair with $J^P = 1^+$ in the $\bar{3}_c$ color state and n=3 defines the $q\bar{q}$-pairs, corresponding to mesons with quantum numbers: $J^{PC} = 0^{++}, 1^{++}, 2^{++}, 0^{-+}, 1^{--}$ and $I_z = 0,1$.

In the case in question the interacting quarks do not produce a bound state; therefore the integration in Eqs.(3) - (6) below is carried out from the threshold $4m^2$ to the cut-off $\Lambda$. The system of integral equations, corresponding to Fig. 1 (the meson state with $J^{PC} = 0^{++}$ and diquark with $J^P = 0^+$), can be described as:

$$A_1(s, s_{1234}, s_{12}, s_{34}) = \frac{\lambda_1 B_3(s_{12}) B_1(s_{34})}{[1 - B_3(s_{12})][1 - B_1(s_{34})]} + 3\hat{J}_2(3,1) A_4(s, s_{1234}, s'_{23}, s'_{234}) +$$
$$+ 2\hat{J}_2(3,1) A_3(s, s_{1234}, s'_{13}, s'_{134}) + 2\hat{J}_1(3) A_3(s, s_{1234}, s'_{15}, s_{125}) + 2\hat{J}_1(3) A_4(s, s_{1234}, s'_{25}, s_{125}) +, \quad (3)$$
$$+ 2\hat{J}_1(1) A_4(s, s_{1234}, s'_{35}, s_{345})$$

$$A_2(s, s_{1234}, s_{25}, s_{34}) = \frac{\lambda_2 B_1(s_{25}) B_1(s_{34})}{[1 - B_1(s_{25})][1 - B_1(s_{34})]} + 6\hat{J}_2(1,1) A_4(s, s_{1234}, s'_{23}, s'_{234}) +$$
$$+ 8\hat{J}_1(1) A_3(s, s_{1234}, s'_{12}, s_{125}) \quad (4)$$

$$A_3(s, s_{1234}, s_{13}, s_{134}) = \frac{\lambda_3 B_3(s_{13})}{1 - B_3(s_{13})} + 8\hat{J}_3(3) A_1(s, s_{1234}, s'_{12}, s'_{34}), \quad (5)$$

$$A_4(s, s_{1234}, s_{23}, s_{234}) = \frac{\lambda_4 B_1(s_{23})}{1 - B_1(s_{23})} + 2\hat{J}_3(1) A_2(s, s_{1234}, s'_{25}, s'_{34}) + 2\hat{J}_3(1) A_1(s, s_{1234}, s'_{12}, s'_{34}), \quad (6)$$

were $\lambda_i$ are the current constants. We introduced the integral operators:

$$\hat{J}_1(l) = \frac{G_l(s_{12})}{[1 - B_l(s_{12})]} \int_{4m^2}^{\Lambda} \frac{ds'_{12}}{\pi} \frac{G_l(s'_{12}) \rho_l(s'_{12})}{s'_{12} - s_{12}} \int_{-1}^{+1} \frac{dz_1}{2}, \quad (7)$$

$$\hat{J}_2(l, p) = \frac{G_l(s_{12}) G_p(s_{34})}{[1 - B_l(s_{12})][1 - B_p(s_{34})]} \times$$
$$\times \int_{4m^2}^{\Lambda} \frac{ds'_{12}}{\pi} \frac{G_l(s'_{12}) \rho_l(s'_{12})}{s'_{12} - s_{12}} \int_{4m^2}^{\Lambda} \frac{ds'_{34}}{\pi} \frac{G_p(s'_{34}) \rho_p(s'_{34})}{s'_{34} - s_{34}} \int_{-1}^{+1} \frac{dz_3}{2} \int_{-1}^{+1} \frac{dz_4}{2}, \quad (8)$$

$$\hat{J}_3(l) = \frac{G_l(s_{12}, \widetilde{\Lambda})}{1 - B_l(s_{12}, \widetilde{\Lambda})} \times$$

$$\times \frac{1}{4\pi} \int_{4m^2}^{\widetilde{\Lambda}} \frac{ds'_{12}}{\pi} \frac{G_l(s'_{12}, \widetilde{\Lambda}) \rho_l(s'_{12})}{s'_{12} - s_{12}} \int_{-1}^{+1} \frac{dz_1}{2} \int_{-1}^{+1} dz \int_{z_2^-}^{z_2^+} dz_2 \frac{1}{\sqrt{1 - z^2 - z_1^2 - z_2^2 + 2zz_1z_2}}, \quad (9)$$

were $l, p$ are equal 1 or 3. If we use the diquark state with $J^P = 1^+$ and the meson with $J^{PC} = 0^{++}, 1^{++}, 2^{++}, 0^{-+}, 1^{--}$, $l, p$ are equal 2 or 3. Here $m$ is a quark mass.

Hereafter we suggest that some unknown (not large) contribution from small distances which might be taken into account by the cut-off procedure. In Eqs.(7) – (9) we choose the "hard" cut-off, but we can use also the "soft" cut-off, for instance $G_n(s_{ik}) = G_n \exp(-(s_{ik} - 4m^2)^2 / \Lambda^2)$. It does not change essentially the calculated mass spectrum.

In Eqs.(7) and (9) $z_1$ is the cosine of the angle between the relative momentum of the particles 1 and 2 in the intermediate state and the momentum of the particle 3 in the final state, is taken in the c.m. of particles 1 and 2. In Eq.(9) $z$ is the cosine of the angle between the momenta of the particles 3 and 4 in the final state, taken in the c.m. of particles 1 and 2. $z_2$ is the cosine of the angle between the relative momentum of particles 1 and 2 in the intermediate state and the momentum of the particle 4 in the final state, taken in the c.m. of particles 1 and 2. In Eq. (8): $z_3$ is the cosine of the angle between relative momentum of particles 1 and 2 in the intermediate state and the relative momentum of particles 3 and 4 in the intermediate state, taken in the c.m. of particles 1 and 2. $z_4$ is the cosine of the angle between the relative momentum of the particles 3 and 4 in the intermediate state and momentum of the particle 1 in the intermediate state, taken in the c.m. of particles 3, 4.

Using the relation of Appendix A we can pass from the integration of the cosines of the angles to the integration of the subenergies.

Let us extract two-particle singularities in the amplitudes $A_1(s, s_{1234}, s_{12}, s_{34})$, $A_2(s, s_{1234}, s_{25}, s_{34})$, $A_3(s, s_{1234}, s_{13}, s_{134})$, and $A_4(s, s_{1234}, s_{23}, s_{234})$:

$$A_1(s, s_{1234}, s_{12}, s_{34}) = \frac{\alpha_1(s, s_{1234}, s_{12}, s_{34}) B_3(s_{12}) B_1(s_{34})}{[1 - B_3(s_{12})][1 - B_1(s_{34})]}. \quad (10)$$

$$A_2(s,s_{1234},s_{25},s_{34}) = \frac{\alpha_2(s,s_{1234},s_{25},s_{34})B_1(s_{25})B_1(s_{34})}{[1-B_1(s_{25})][1-B_1(s_{34})]}. \tag{11}$$

$$A_3(s,s_{1234},s_{13},s_{134}) = \frac{\alpha_3(s,s_{1234},s_{13},s_{134})B_3(s_{13})}{1-B_3(s_{13})}, \tag{12}$$

$$A_4(s,s_{1234},s_{23},s_{234}) = \frac{\alpha_4(s,s_{1234},s_{23},s_{234})B_1(s_{23})}{1-B_1(s_{23})}, \tag{13}$$

We do not extract three- and four-particle singularities, because they are weaker than two-particle singularities.

We used the classification of singularities, which was proposed in paper [20] for the two and three particle singularities. The construction of approximate solution of Eqs.(3) - (6) is based on the extraction of the leading singularities of the amplitudes. The main singularities in $s_{ik} \approx 4m^2$ are from pair rescattering of the particles i and k. First of all there are threshold square-root singularities. Also possible singularities are pole singularities which correspond to the bound states. The diagrams of Fig.1 apart from two-particle singularities have triangular singularities and singularities defining the interaction of four and five particles. Such classification allowed us to find the corresponding solutions of Eqs.(3) - (6) by taking into account some definite number of leading singularities and neglecting all the weaker ones. We considered the approximation which defines two-particle, triangle, four- and five-particle singularities. The functions $\alpha_1(s,s_{1234},s_{12},s_{34})$, $\alpha_2(s,s_{1234},s_{25},s_{34})$, $\alpha_3(s,s_{1234},s_{13},s_{134})$, and $\alpha_4(s,s_{1234},s_{23},s_{234})$ are smooth functions of $s_{ik}$, $s_{ijk}$, $s_{ijkl}$, $s$ as compared with the singular part of the amplitudes, hence they can be expanded in a series in the singularity point and only the first term of this series should be employed further. Using this classification one define the reduced amplitudes $\alpha_1$, $\alpha_2$, $\alpha_3$, $\alpha_4$ as well as the B-functions in the middle point of the physical region of Dalitz-plot at the point $s_0$:

$$s_0^{ik} = s_0 = \frac{s+15m^2}{10}, \tag{14}$$

$$s_{123} = 3s_0 - 3m^2, \quad s_{1234} = 6s_0 - 8m^2.$$

Such a choice of point $s_0$ allows to replace the integral equations (3) - (6) (Fig. 1) by the algebraic equations (15) - (18) respectively:

$$\alpha_1 = \lambda_1 + 3\alpha_4 J_2(3,1,1) + 2\alpha_3 J_2(3,1,3) + 2\alpha_3 J_1(3,3) + 2\alpha_4 J_1(3,1) + 2\alpha_4 J_1(1,1), \tag{15}$$

$$\alpha_2 = \lambda_2 + 6\alpha_4 J_2(1,1,1) + 8\alpha_3 J_1(1,3), \tag{16}$$

$$\alpha_3 = \lambda_3 + 8\alpha_1 J_3(3,3,1), \tag{17}$$

$$\alpha_4 = \lambda_4 + 2\alpha_2 J_3(1,1,1) + 2\alpha_1 J_3(1,1,3). \tag{18}$$

We use the functions $J_1(l,p)$, $J_2(l,p,r)$, $J_3(l,p,r)$ ($l,p,r = 1, 2, 3$):

$$J_1(l,p) = \frac{G_l^2(s_0^{12})B_p(s_0^{15})}{B_l(s_0^{12})} \int_{4m^2}^{\Lambda} \frac{ds_{12}'}{\pi} \frac{\rho_l(s_{12}')}{s_{12}' - s_0^{12}} \int_{-1}^{+1} \frac{dz_1}{2} \frac{1}{1 - B_p(s_{15}')}, \tag{19}$$

$$J_2(l,p,r) = \frac{G_l^2(s_0^{12})G_p^2(s_0^{34})B_r(s_0^{13})}{B_l(s_0^{12})B_p(s_0^{34})} \times$$

$$\times \int_{4m^2}^{\Lambda} \frac{ds_{12}'}{\pi} \frac{\rho_l(s_{12}')}{s_{12}' - s_0^{12}} \int_{4m^2}^{\Lambda} \frac{ds_{34}'}{\pi} \frac{\rho_p(s_{34}')}{s_{34}' - s_0^{34}} \int_{-1}^{+1} \frac{dz_3}{2} \int_{-1}^{+1} \frac{dz_4}{2} \frac{1}{1 - B_r(s_{13}')} \tag{20}$$

$$J_3(l,p,r) = \frac{G_l^2(s_0^{12},\tilde{\Lambda})B_p(s_0^{13})B_r(s_0^{24})}{1 - B_l(s_0^{12},\tilde{\Lambda})} \frac{1 - B_l(s_0^{12})}{B_l(s_0^{12})} \times$$

$$\times \frac{1}{4\pi} \int_{4m^2}^{\tilde{\Lambda}} \frac{ds_{12}'}{\pi} \frac{\rho_l(s_{12}')}{s_{12}' - s_0^{12}} \int_{-1}^{+1} \frac{dz_1}{2} \int_{-1}^{+1} dz \int_{z_2^-}^{z_2^+} dz_2 \frac{1}{\sqrt{1 - z^2 - z_1^2 - z_2^2 + 2zz_1z_2}} \frac{1}{[1 - B_p(s_{13}')][1 - B_r(s_{24}')]} \tag{21}$$

The other choices of point $s_0$ do not change essentially the contributions of $\alpha_1$, $\alpha_2$, $\alpha_3$ and $\alpha_4$; therefore we omit the indexes $s_0^{ik}$. Since the vertex functions depend only slightly on energy it is possible to treat them as constants in our approximation and determine them in a way similar to that used in [21, 22].

The integration contours of functions $J_1$, $J_2$, $J_3$ are given in the Appendix B (Figs. 6, 7, 8). The equations, which are similar to Eqs.(15) – (18) but correspond to other low-lying mesons with isospin $I_z = 0,1$, $J^{PC} = 0^{++}, 1^{++}, 2^{++}, 0^{-+}, 1^{--}$ and diquarks with $J^P = 0^+, 1^+$ (graphic equations Fig.1 - 5).

The solutions of the system of equations are considered as:

$$\alpha_i(s) = F_i(s,\lambda_i)/D(s), \tag{22}$$

where zeros of $D(s)$ determinants define the masses of bound states of pentaquarks. $F_i(s,\lambda_i)$ are the functions of $s$ and $\lambda_i$. The functions $F_i(s,\lambda_i)$ determine the contributions of subamplitudes to the pentaquark amplitude.

III. Calculation results.

The poles of the reduced amplitudes $\alpha_1$, $\alpha_2$, $\alpha_3$, $\alpha_4$ correspond to the bound states and determine the masses of $N$ and $\Delta$ – isobar pentaquarks. In our calculation we assume a quark mass $m = 410\, MeV$ ($m \geq \frac{1}{5} m_{\frac{5}{2}^+}(2000)$). The model has only three parameters similar to our model of hybrid baryons [11]. The gluon coupling constant $g = 0.417$ is fitted by fixing $N$ pentaquark mass $m_{\frac{5}{2}^+}(2000)$. The cut-off parameters $\Lambda_{0^+} = 16.5$ and $\Lambda_{1^+} = 20.12$ can be determined by the nucleon (1710) and $\Delta$ – isobar (1950) pentaquarks respectively. The calculated mass values of low-lying nucleon and $\Delta$ – isobar pentaquarks are shown in Tables 1 and 2. We found the lowest masses of $N$ pentaquarks with $J^P = \frac{1}{2}^-$ M=1583 MeV, $J^P = \frac{1}{2}^+$ M=1686 MeV and $\Delta$ – isobar pentaquarks with $J^P = \frac{3}{2}^-$ M=1453 MeV, $J^P = \frac{3}{2}^+$ M=1485 MeV. If we increase the quark mass, the masses of the lowest $\Delta$ – isobar pentaquarks can be increased, but the masses of the pentaquarks will be largest of the calculated masses (Tables 1, 2). The low-lying $\Delta$ – isobar pentaquark masses are smaller than the $N$ pentaquark masses. It depends on the different interactions in the diquark channels $J^P = 0^+, 1^+$. The calculated values of the pentaquark masses are compared to the experimental data [12]. We predict the degeneracy of some states. The calculation of pentaquark amplitude estimates the contributions of four subamplitudes. The main contributions to the pentaquark amplitude are determined by the subamplitudes, which include the low-lying meson states. Tables 1, 2 show the contributions of the following subamplitudes: $A_1$ ($BM$), $A_2$ ($D\bar{q}D$), $A_3$ ($Mqqq$), $A_4$ ($Dqq\bar{q}$). We found that the contributions of $A_1$ and $A_3$ subamplitudes are about 30-70 % of the pentaquark contribution.

Williams and Gueye have developed a quark molecular model (QMM) of strangeness S=0 Pentaquark baryons assuming molecule structure and predicted the low-lying mass spectrum [23]. They considered the molecules $uds - u\bar{s}$ and received the masses about 1800-2000 MeV. We take into account only u,d-pentaquarks with S=0: $udud\bar{u}, udud\bar{d}, uuuu\bar{u}, dddd\bar{d}$, and received smaller masses of low-lying pentaquarks.

IV Conclusion.

In strongly bound system of light quarks such as the baryons consideration, where $p/m \approx 1$ the approximation of nonrelativistic kinematics and dynamics not justified.

In our relativistic five-quark model (Faddeev – Yakubovsky type approach) we calculated the masses of low-lying pentaquarks. We used only the u, d quarks. The quark amplitudes obey the global color symmetry. The masses of the constituent quarks are equal to 410 MeV. We considered the scattering amplitudes of the constituent quarks. The poles of these amplitudes determine the masses of low-lying pentaquarks. The derived five-quark amplitude consists of four subamplitudes: $BM$, $D\bar{q}D$, $Mqqq$, $Dqq\bar{q}$, where $B$, $M$ and $D$ are the baryon, meson and the diquark respectively.

Unlike mesons, all half-integral spin and parity quantum numbers are allowed in the baryon sector, so that experiments search for such pentaquark states are not simple. Furthermore, no decay channels are a priori forbidden. These two facts make identification of a pentaquark difficult.

We have been treating the quarks as real particles. However, in the soft region the quark diagrams should be treated as spectral integrals of the quark mass with spectral density $\rho(m^2)$: the integration of the quark mass in the amplitudes eliminates the quark singularities and introduces the hadron ones. One can believe that the approximation:

$$\rho(m^2) \Rightarrow \delta(m^2 - m_q^2) \qquad (20)$$

could be possible for the low-lying hadrons (here $m_q$ is the "mass" of the constituent quark). We hope that the approach given by Eq. (20) is sufficiently good for the calculation of the low-lying pentaquarks being carried out here.

Acknowledgments.

One of the authors (S.M. Gerasyuta) is indebted to the Institut fur Kernphysik Forschungzentrum Julich for hospitality where this work was initiated. The authors would like to thank T. Barnes, D.I. Diakonov, S. Krewald, N.N. Nikolaev, P.R. Page for useful discussions. This research was supported in part by the Russian Ministry of Education, Program "Universities of Russia" under Contract № 01.20.00. 06448.

APPENDIX A

We can go over from the integration with respect of the cosines of angles to the integration with respect to the energy variables by using the relations:

$$s'_{13} = 2m^2 + \frac{s_{123} - s'_{12} - m^2}{2} + \frac{z_1}{2}\sqrt{\frac{s'_{12} - 4m^2}{s'_{12}}[(s_{123} - s'_{12} - m^2)^2 - 4s'_{12}m^2]}, \quad (A1)$$

$$s'_{24} = 2m^2 + \frac{s_{124} - s'_{12} - m^2}{2} + \frac{z_1}{2}\sqrt{\frac{s'_{12} - 4m^2}{s'_{12}}[(s_{124} - s'_{12} - m^2)^2 - 4s'_{12}m^2]}, \quad (A2)$$

$$z = \frac{2s'_{12}(s_{1234} + s'_{12} - s_{123} - s'_{124}) - (s_{123} - s'_{12} - m^2)(s'_{124} - s'_{12} - m^2)}{\sqrt{[(s_{123} - s'_{12} - m^2)^2 - 4m^2 s'_{12}][(s'_{124} - s'_{12} - m^2)^2 - 4m^2 s'_{12}]}}. \quad (A3)$$

$$s'_{134} = m^2 + s'_{34} + \frac{s_{1234} - s'_{12} - s'_{34}}{2} + \frac{z_3}{2}\sqrt{\frac{s'_{12} - 4m^2}{s'_{12}}[(s_{1234} - s'_{12} - s'_{34})^2 - 4s'_{12}s'_{34}]} \quad (A4)$$

$$s'_{13} = 2m^2 + \frac{s'_{134} - s'_{34} - m^2}{2} + \frac{z_4}{2}\sqrt{\frac{s'_{34} - 4m^2}{s'_{34}}\left[(s'_{134} - s'_{34} - m^2)^2 - 4m^2 s'_{34}\right]} \quad (A5)$$

The integration under consideration is taken in the physical region, where $-1 \leq z_i \leq 1$ ($i = 1, 2, 3, 4$) and one can define the integration region for the invariant variables. Therefore for $s'_{124}$ we have condition $0 \leq z^2 \leq 1$ and

$$s'^{\pm}_{124} = s'_{12} + m^2 + \frac{(s_{1234} - s_{123} - m^2)(s_{123} + s'_{12} - m^2)}{2s_{123}} \pm$$
$$\pm \frac{1}{2s_{123}}\sqrt{[(s_{123} - s'_{12} - m^2)^2 - 4m^2 s'_{12}][(s_{1234} - s_{123} - m^2)^2 - 4m^2 s_{123}]}, \quad (A6)$$

and the region of integration on $s'_{12}$ in $J_3$:

$$\tilde{\Lambda} = \begin{cases} \Lambda, & \text{if } \Lambda \leq (\sqrt{s_{123}} + m)^2 \\ (\sqrt{s_{123}} + m)^2, & \text{if } \Lambda > (\sqrt{s_{123}} + m)^2 \end{cases}. \quad (A7)$$

The integration contour 1 (Fig. 6) corresponds to the relation $s_{123} < (\sqrt{s_{12}} - m)^2$; the contour 2 is defined by the relation $(\sqrt{s_{12}} - m)^2 < s_{123} < (\sqrt{s_{12}} + m)^2$. The point $s_{123} = (\sqrt{s_{12}} - m)^2$ is not singular, and the contour around this point at $s_{123} + i\varepsilon$ and $s_{123} - i\varepsilon$ gives identical result. $s_{123} = (\sqrt{s_{12}} + m)^2$ is the singular point, but in our case the integration contour can not pass through this point because the region in consideration is situated below the production threshold of the four particles $s_{1234} < 16m^2$. A similar situation occurs for the integration over $s_{13}$ in the function $J_3$. In Fig. 6, 7b, 8 the dotted lines define the square-root cut of the Chew-Mandelstam functions. They correspond to two-particles threshold and also three-particles threshold in Fig. 7a. The integration contour 1 (Fig. 7a) is determined by $s_{1234} < (\sqrt{s_{12}} - \sqrt{s_{34}})^2$, the contour 2 corresponds to the case $(\sqrt{s_{12}} - \sqrt{s_{34}})^2 < s_{1234} < (\sqrt{s_{12}} + \sqrt{s_{34}})^2$. $s_{1234} = (\sqrt{s_{12}} - \sqrt{s_{34}})^2$ is not a singular point, and the contour around this point at $s_{1234} + i\varepsilon$ and $s_{1234} - i\varepsilon$ gives identical result. The integration contour 1 (Fig. 7b) is determined by the region $s_{1234} < (\sqrt{s_{12}} - \sqrt{s_{34}})^2$ and $s_{134} < (\sqrt{s_{34}} - m)^2$, the integration contour 2 corresponds to $s_{1234} < (\sqrt{s_{12}} - \sqrt{s_{34}})^2$ and $(\sqrt{s_{34}} - m)^2 \le s_{134} < (\sqrt{s_{34}} + m)^2$. The contour 3 is defined by $(\sqrt{s_{12}} - \sqrt{s_{34}})^2 < s_{1234} < (\sqrt{s_{12}} + \sqrt{s_{34}})^2$. Here the singular point would be $s_{134} = (\sqrt{s_{34}} + m)^2$. But in our case this point is not reachable, if one has the condition $s_{1234} < 16m^2$. We have to consider the integration over $s_{24}$ in the function $J_3$. While $s_{124} < s_{12} + 5m^2$ the integration is carried out along the complex axis (the contour 1, Fig. 8). If we come to the point $s_{124} = s_{12} + 5m^2$, then the contour reach the square-root cut of Chew-Mandelstam function (contour 2, Fig. 8). In this case the part of the integration contour in the nonphysical region is along the real axis. The other part of integration contour along the complex axis corresponds to physical region. The suggested calculation shows that the contribution of the integration over the nonphysical region is small [21, 22].

Table I. Low-lying nucleon pentaquark masses and contributions of subamplitudes $BM$, $D\bar{q}D$, $Mqqq$ and $Dqq\bar{q}$ to pentaquark amplitude in percentage of probability (diquark with $J^P = 0^+$).

| Fig. № | Meson $J^{PC}$ | $J^P$ | Mass, MeV | $A_1$ ($BM$) | $A_2$ ($D\bar{q}D$) | $A_3$ ($Mqqq$) | $A_4$ ($Dqq\bar{q}$) |
|---|---|---|---|---|---|---|---|
| 1 | $0^{++}$ | $\frac{1}{2}^+$ | 1686(1440) | 32.01 | 16.57 | 37.29 | 14.13 |
| 1 | $1^{++}$ | $\frac{1}{2}^+$ | 1710(1710) | 28.50 | 20.82 | 38.39 | 12.29 |
| 2 | $1^{++}$ | $\frac{3}{2}^+$ | 1765(1720) | 34.19 | - | 51.82 | 13.99 |
| 2 | $2^{++}$ | $\frac{3}{2}^+$ | 1922(1900) | 30.88 | - | 55.46 | 13.66 |
| 3 | $2^{++}$ | $\frac{5}{2}^+$ | 2000(2000) | 30.75 | - | 69.25 | - |
| 3 | $0^{-+}$ | $\frac{1}{2}^-$ | 1583(1535) | 41.65 | - | 58.35 | - |
| 3 | $1^{--}$ | $\frac{1}{2}^-, \frac{3}{2}^-$ | 1973(1700) | 31.51 | - | 68.49 | - |

Parameters of model: quark mass $m$ = 410 MeV, cut-off parameter $\Lambda$ =16,5; gluon coupling constant $g$ =0.417. Experimental mass values of nucleon pentaquark are given in parentheses [12].

Table II. Low-lying $\Delta$ - isobar pentaquark masses and contributions of subamplitudes $BM$, $D\bar{q}D$, $Mqqq$ and $Dqq\bar{q}$ to pentaquark amplitude in percentage of probability (diquark with $J^P = 1^+$).

| Fig. № | Meson $J^{PC}$ | $J^P$ | Mass, MeV | $A_1$ ($BM$) | $A_2$ ($D\bar{q}D$) | $A_3$ ($Mqqq$) | $A_4$ ($Dqq\bar{q}$) |
|---|---|---|---|---|---|---|---|
| 4 | $0^{++}$ | $\frac{1}{2}^+, \frac{3}{2}^+$ | 1485(1600) | 31.60 | 6.42 | 33.93 | 28.05 |
| 4 | $1^{++}$ | $\frac{1}{2}^+, \frac{3}{2}^+, \frac{5}{2}^+$ | 1550(1750) | 28.08 | 8.88 | 42.09 | 20.95 |
| 4 | $2^{++}$ | $\frac{1}{2}^+, \frac{3}{2}^+, \frac{5}{2}^+$ | 1736(1920) | 24.53 | 13.25 | 44.07 | 18.15 |
| 5 | $2^{++}$ | $\frac{7}{2}^+$ | 1950(1950) | 24.99 | - | 75.01 | - |
| 5 | $0^{-+}$ | $\frac{1}{2}^-$ | 1453(1620) | 38.13 | - | 61.87 | - |
| 5 | $1^{--}$ | $\frac{1}{2}^-, \frac{3}{2}^-$ | 1920(1940) | 25.97 | - | 74.03 | - |

Parameters of model: quark mass $m$ = 410 MeV, cut-off parameter $\Lambda$ =20,1; gluon constant $g$ =0.417. Experimental mass values of $\Delta$- isobar pentaquarks are given in parentheses [12].

Table III. Vertex functions

| $J^{PC}$ | $G_n^2$ |
|---|---|
| $0^+$ (n=1) | $4g/3 - 8gm^2/(3s_{ik})$ |
| $1^+$ (n=2) | $2g/3$ |
| $0^{-+}$ (n=3) | $8g/3 - 16gm^2/(3s_{ik})$ |
| $1^{--}$ (n=3) | $4g/3$ |
| $0^{++}$ (n=3) | $8g/3$ |
| $1^{++}$ (n=3) | $4g/3$ |
| $2^{++}$ (n=3) | $4g/3$ |

Table IV. Coefficient of Chew-Mandelstam functions for n = 3 (meson states) and diquarks n = 1 ($J^P = 0^+$), n = 2 ($J^P = 1^+$).

| $J^{PC}$ | n | $\alpha(J^{PC},n)$ | $\beta(J^{PC},n)$ |
|---|---|---|---|
| $0^{++}$ | 3 | 1/2 | -1/2 |
| $1^{++}$ | 3 | 1/2 | 0 |
| $2^{++}$ | 3 | 3/10 | 1/5 |
| $0^{-+}$ | 3 | 1/2 | 0 |
| $1^{--}$ | 3 | 1/3 | 1/6 |
| $0^+$ | 1 | 1/2 | 0 |
| $1^+$ | 2 | 1/3 | 1/6 |

Figure captions.

Fig.1. Graphic representation of the equations for the five-quark subamplitudes $A_1(s,s_{1234},s_{12},s_{34})$ ($BM$), $A_2(s,s_{1234},s_{25},s_{34})$ ($D\bar{q}D$), $A_3(s,s_{1234},s_{13},s_{134})$ ($Mqqq$), and $A_4(s,s_{1234},s_{23},s_{234})$ ($Dqq\bar{q}$) using the low-lying mesons with $J^{PC}=0^{++},1^{++},2^{++},0^{-+},1^{--}$ and diquark with $J^P=0^+$.

Fig.2. Graphic representation of the equations for the five-quark subamplitudes $A_1(s,s_{1234},s_{12},s_{34})$ ($BM$), $A_3(s,s_{1234},s_{13},s_{134})$ ($Mqqq$), and $A_4(s,s_{1234},s_{23},s_{234})$ ($Dqq\bar{q}$) using the low-lying mesons with $J^{PC}=0^{++},1^{++},2^{++},0^{-+},1^{--}$ and diquark with $J^P=0^+$.

Fig.3. Graphic representation of the equations for the five-quark subamplitudes $A_1(s,s_{1234},s_{12},s_{34})$ ($BM$) and $A_3(s,s_{1234},s_{13},s_{134})$ ($Mqqq$) using the low-lying mesons with $J^{PC}=0^{++},1^{++},2^{++},0^{-+},1^{--}$ and diquark with $J^P=0^+$.

Fig.4. Graphic representation of the equations for the five-quark subamplitudes $A_1(s,s_{1234},s_{12},s_{34})$ ($BM$), $A_2(s,s_{1234},s_{25},s_{34})$ ($D\bar{q}D$), $A_3(s,s_{1234},s_{13},s_{134})$ ($Mqqq$), and $A_4(s,s_{1234},s_{23},s_{234})$ ($Dqq\bar{q}$) using the low-lying mesons with $J^{PC}=0^{++},1^{++},2^{++},0^{-+},1^{--}$ and diquark with $J^P=1^+$.

Fig.5. Graphic representation of the equations for the five-quark subamplitudes $A_1(s,s_{1234},s_{12},s_{34})$ ($BM$) and $A_3(s,s_{1234},s_{13},s_{134})$ ($Mqqq$) using the low-lying mesons with $J^{PC}=0^{++},1^{++},2^{++},0^{-+},1^{--}$ and diquark with $J^P=1^+$.

Fig. 6. Contours of integration 1, 2 in the complex plane $s_{13}$ for the functions $J_1$, $J_3$.

Fig. 7. Contours of integration 1, 2, 3 in the complex plane $s_{134}$ (a) and $s_{13}$ (b) for the function $J_2$.

Fig. 8. Contours of integration 1, 2 in the complex plane $s_{24}$ for the function $J_3$.

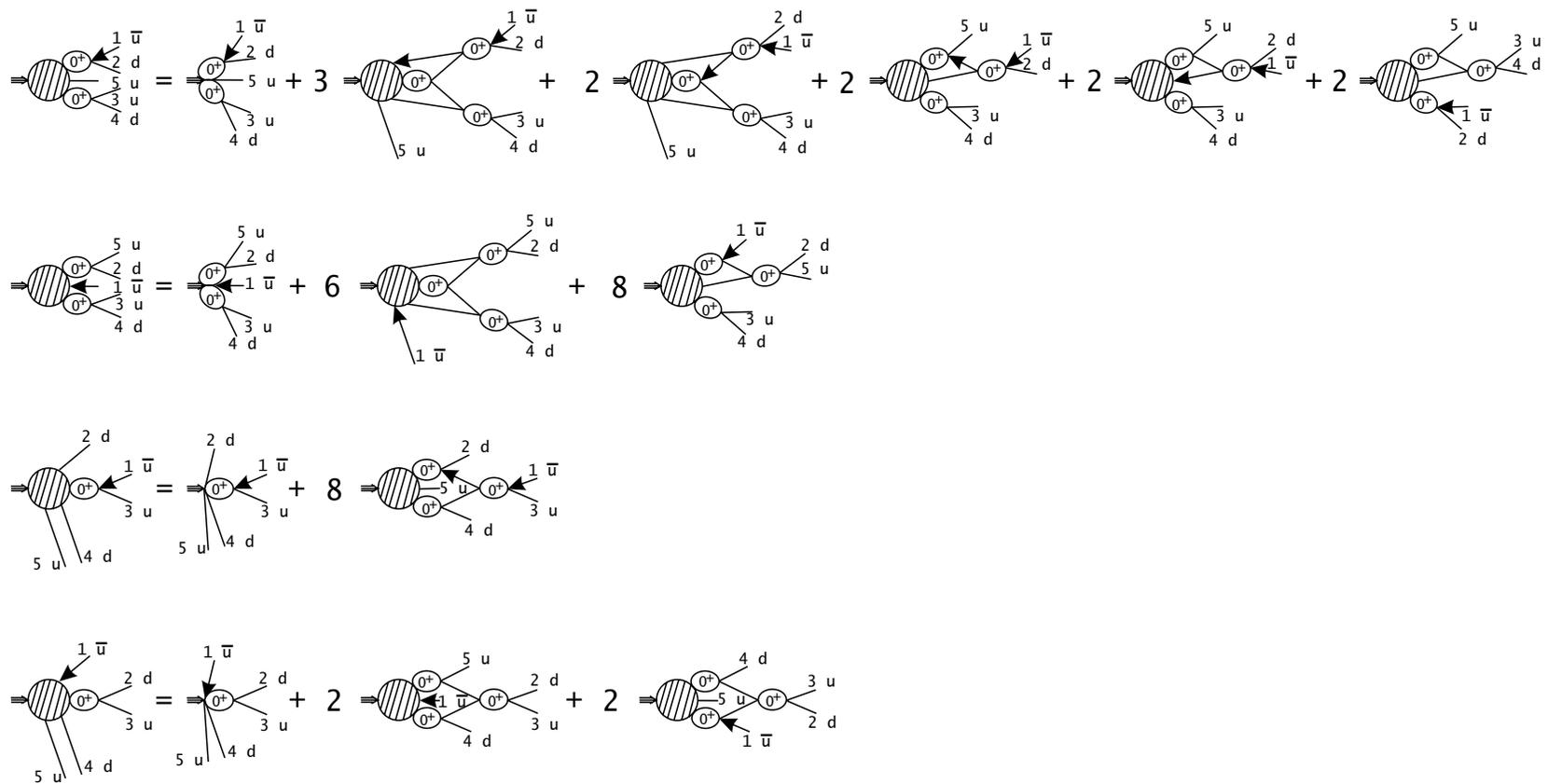

Fig. 1

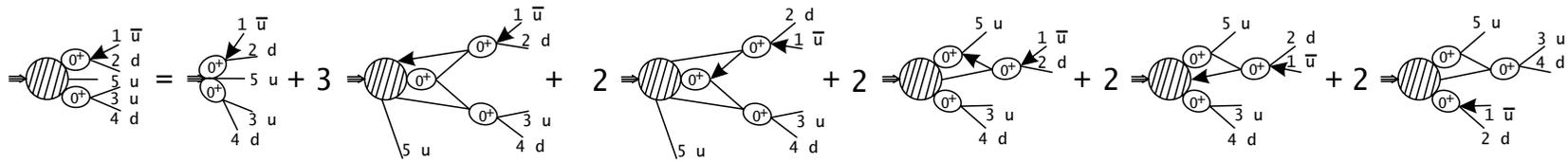

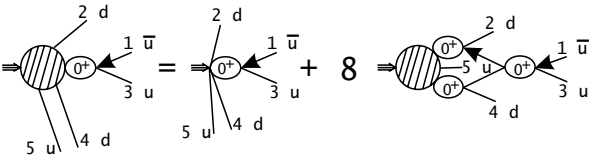

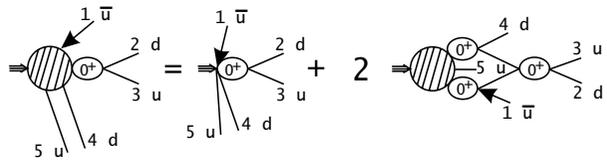

Fig. 2

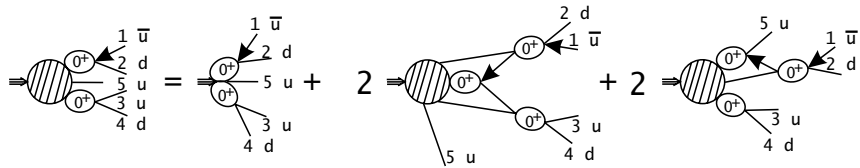

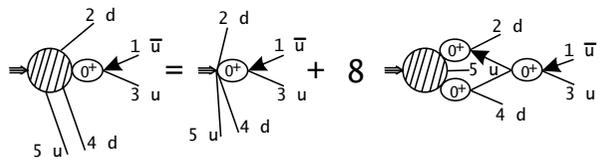

Fig. 3

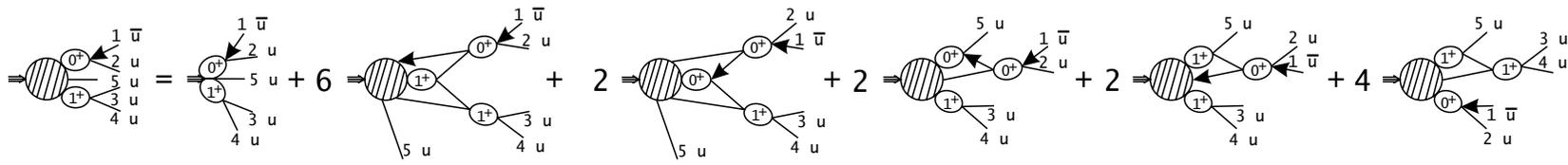

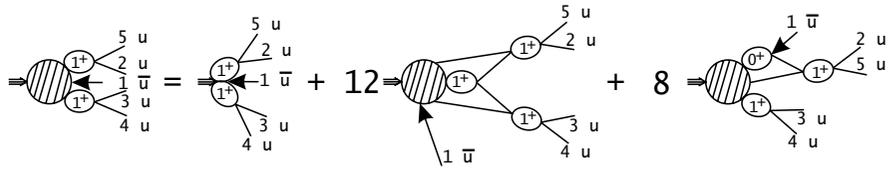

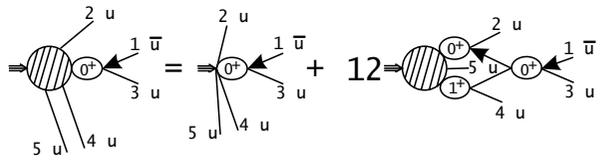

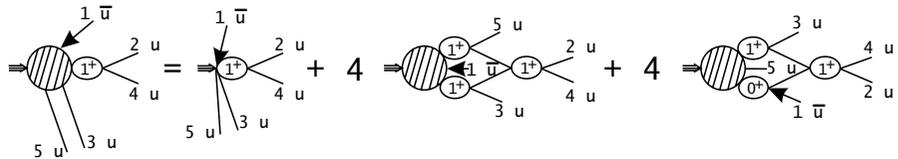

Fig. 4

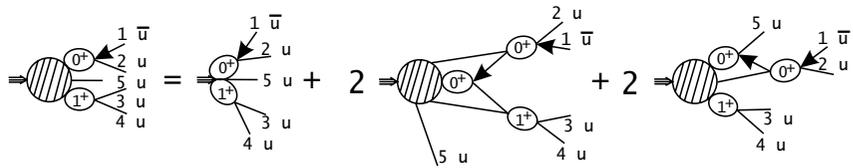

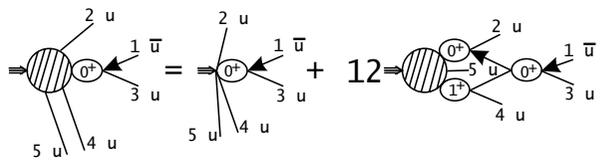

Fig. 5

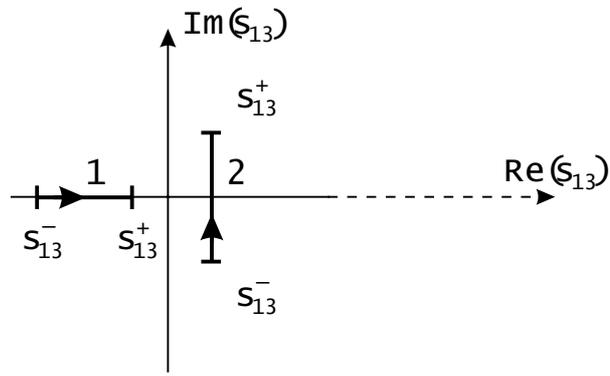

Fig. 6

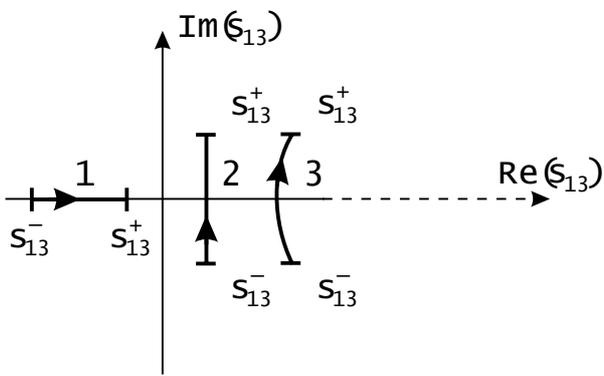

Fig. 7b

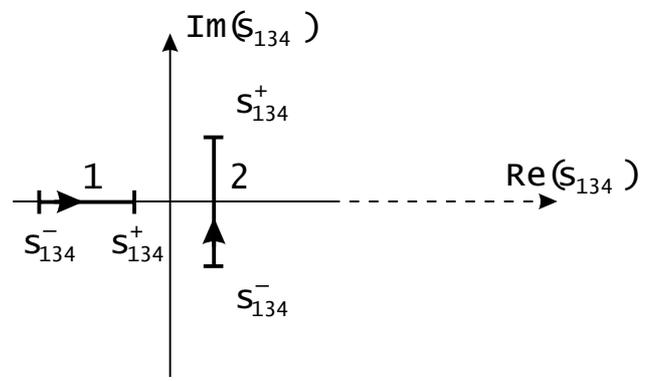

Fig. 7a

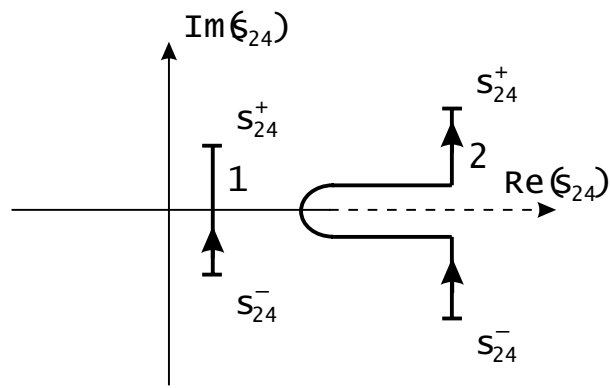

Fig. 8